\documentclass[twocolumn,superscriptaddress,showpacs,floatfix]{revtex4}
\usepackage{graphicx} 
\begin{document}   

\title{Knots in Charged Polymers} 
\author{Paul G. Dommersnes}
\email{paul.dommersnes@phys.ntnu.no}
\affiliation{Department of Physics, Massachusetts
Institute of Technology, Cambridge, Massachusetts 02139}
\affiliation{Department of Physics, Norwegian University of Science and Technology, 7491 Trondheim, Norway}
\author{Yacov Kantor}
\email{kantor@post.tau.ac.il}
\affiliation{School for Physics and Astronomy, Tel Aviv
University, Tel Aviv 69978, Israel}
\affiliation{Department of Physics, Massachusetts
Institute of Technology, Cambridge, Massachusetts 02139}
\author{Mehran Kardar}
\affiliation{Department of Physics, Massachusetts
Institute of Technology, Cambridge,
Massachusetts 02139}

\date{\today} 

\begin{abstract}
The interplay of topological constraints and  Coulomb interactions in 
static and dynamic properties of charged polymers 
is investigated by  numerical simulations and scaling arguments. 
In the absence of screening, the long-range interaction localizes irreducible
topological constraints into tight molecular knots,
while composite constraints are factored and separated.
Even when the forces are screened, tight knots may survive as
local (or even global) equilibria, as long as the overall rigidity 
of the polymer is dominated by the Coulomb interactions.
As entanglements involving tight knots are not easy to eliminate,
their presence greatly influences the relaxation times of the system.
In particular, we find that tight knots in open polymers are removed
by diffusion along the chain, rather than by opening up.
The knot diffusion coefficient actually decreases with its charge density, 
and for highly charged polymers the knot's position appears frozen.  
\end{abstract}
\widetext
\maketitle
\section{Introduction}
A polymer chain can be easily deformed, but since it cannot cross itself,
it is subject to topological constraints.
These constraints can be temporary, such as entanglements
between linear polymers, or permanent if the chains are closed (ring
polymers) or cross-linked.  
Understanding the influence of  topological entanglements on
static and dynamic properties of polymers is a long-standing
issue~\cite{degennesSC,edwards}, which has recently found
renewed interest in the context of {\it knotted biopolymers}.  
DNA in the cell can change its
topology by the {\it topoisomerase} enzymes that pass one strand
through another, in the process either creating or removing knots~\cite{topo}.
Synthetic RNA trefoil knots have been used to prove the existence of a 
similar (previously unknown) topology changing enzyme~\cite{wang}.  
There is also much interest in developing artificial biopolymers,
for example as molecular building blocks or for DNA-based computing, 
and in this quest complex knots and links have been created in 
both {\em single} and {\em double} stranded DNA~\cite{seeman}. 
Tight knots have been tied in single molecule experiments on both DNA
and actin filaments  using optical tweezers~\cite{arai}.  

Several theoretical approaches have addressed the influence of topological
constraints in polymer networks and solutions.
In particular, the {\em tube} model~\cite{edwards} in which the constraints are
replaced by a hard confining tube, is quite successful in predicting
relaxation dynamics of polymeric solutions.
In a complementary approach, topological constraints are described
in terms of {\it localized} entanglements or knots, 
that perform collective motions along the polymers~\cite{iwata}. 
Single molecule experiments are now able to probe polymers of
specified topology, and to examine the influence of  knot complexity on basic 
physical properties such as the radius of gyration $R_{\rm g}$. 
A simple scaling picture~\cite{quake} suggests that $R_{\rm g}$ is
reduced as a power of the knot complexity, measured by the minimal number of 
crossings in a projection.
Indeed, a Flory mean field theory of knotted ring polymers~\cite{grosberg,rabin}
incorporating this knot invariant  predicts various scaling dependences on 
knot complexity. 
A topological localization effect is also suggested, in which knots segregate 
in a single relatively compact domain while the rest of the polymer ring expels
all the entanglements and swells freely. 
Recent Monte Carlo simulations in 
Refs.~\cite{guiter,katrich,orlandini,farago} support 
the idea that entropic factors localize topological constraints. 
This is bolstered by analytical arguments on slip-linked polymers~\cite{metz},
and experiments on vibrated granular chains~\cite{Ben-Naim}.

Many biopolymers are highly charged. 
The effect of electrostatics on knotting probability of double stranded DNA 
has been studied in the case where the screening length is smaller than 
the persistence length of the polymer. The effect of the Coulomb interactions 
is then to renormalize the effective thickness of the polymer~\cite{rybenkov-shaw,tesi}.  
However synthetic polymers and single stranded DNA both have an 
intrinsic persistence length of the 
order $\ell_{\rm p}\sim 1$nm~\cite{tinland} which could be
small compared to the electrostatic screening lengths. 
In this paper we explore the influence of topological constraints on
charged polymers in cases  where the screening length is large or
comparable to the intrinsic persistence length. 
In Sec.~\ref{ideal} we start by considering the idealized case of unscreened
Coulomb interactions. This case demonstrates that under long-range
interactions the topological constraints are pulled into tight knots.
As discussed in Sec.~\ref{real}, this conclusion has to be re-examined in
real systems due to finite rigidity of the polymer, thermal fluctuations,
and, most importantly, finite screening.
Surprisingly, we find that tight knots are rather resilient:
They remain as global equilibrium solutions as long as the overall
shape of the polymer is dominated by the (screened) Coulomb interactions.
Tight knots can also remain as metastable states for shorter screening
lengths, as long as the electrostatic bending rigidity is larger than 
the intrinsic one.
Such long-lived tight knots have strong influence on the relaxation dynamics
of the polymers as discussed in Sec.~\ref{dynamics}.
In particular, we find that the most likely way for eliminating topological
entanglements is by diffusion of tight knots along the chain; 
interestingly stronger Coulomb interactions lead to tighter knots that are less mobile.

\section{Unscreened interactions}\label{ideal}

We first consider a simple model of a charged polymer in which
monomers repel each other via {\em unscreened} Coulomb interactions.
The interaction between two charges $e$, in a solvent with dielectric 
constant $\varepsilon$, separated by distance $r$ is $e^2/\varepsilon r$, 
and consequently the overall electrostatic energy of a polymer 
of $N$ monomers is
$V_c=(e^2/\varepsilon)\sum_{i>j}^{N}1/|{\bf r}_i-{\bf r}_j|$, where
${\bf r}_i$ is the position of $i$-th monomer.  
Given a typical separation between adjacent monomers of $a$, 
it is convenient to
introduce the energy scale $\epsilon_o\equiv e^2/\varepsilon a$.  
Initially, we focus on configurations in which the monomers are locally stretched to
form smooth straight segments, gradually curving at a larger length scale
$R$  set by the overall shape. 
For such configurations the Coulomb energy has the form
\begin{equation}
\label{Ec}
E_c(N)=\epsilon_o \left[N\ln\left(\frac{R}{a}\right)+c\,\frac{aN^2}{R}\right]\ ,
\end{equation}
where $c$ is a numerical constant of order unity, and we note the following:
\begin{itemize}
  \item For any smooth curve, the integral of the 
$1/r$--potential leads to a logarithmic
divergence, and consequently the energy of the polymer is {\em overextensive},
and, consequently, the tension on the polymer increases as $\ln N$. 
Therefore,
thermal fluctuations are irrelevant for a sufficiently long polymer, 
whose shape is determined by minimizing the energy.
  \item The second term in Eq.~\ref{Ec} can be regarded as the Coulomb interaction 
between charges (or order $N$) on remote parts of the polymer (distances of order $R$).
Since typically $R\propto Na$, the partition of the energy between the two parts is
not precise, and can be changed by redefining $R$.
  \item The Coulomb interaction prefers to keep the charges far apart, and the
polymer minimizes its energy by assuming a shape with maximal $R$. 
Thus, open polymers simply form straight lines,
while unknotted ring-polymers  form  circles.
\end{itemize}

%%%%%%%%%%%%%%%%%%%%%%%%%%%%%%%%%%%%%%%%%%%%%%%%%%%%%%%%%%%%%%%%%%%%%
\begin{figure}
\begin{center}
\includegraphics[width=8cm]{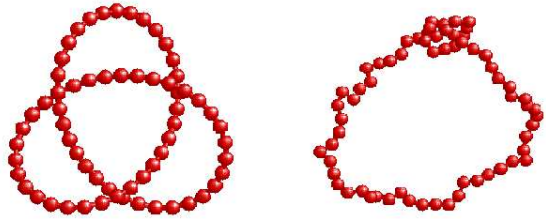}
\caption{\label{trefoil_tight}The initial (left) and equilibrium (right) 
conformations of a 64-monomer charged polymer, at 
$\tilde{T}=1.4$, forming a trefoil knot.
(The right figure is reduced by a factor of 2). }
\end{center}
\end{figure} 
%%%%%%%%%%%%%%%%%%%%%%%%%%%%%%%%%%%%%%%%%%%%%%%%%%%%%%%%%%%%%%%%%%%%%

The above argument can be misleading in the case of a knotted polymer,
as illustrated in Fig.~\ref{trefoil_tight}.
Here, we used Monte Carlo (MC) off-lattice simulations to determine the 
shape of knotted polymers at finite temperature.
Our model polymer consisted of hard sphere monomers connected by 
``tethers''\cite{KKN} that have no energy but limit the distance of a 
connected pair to 1.05 of the hard sphere diameter $a$. 
Fig.~\ref{trefoil_tight} depicts the results of a simulation for a trefoil knot: 
As an initial conformation (left) in this simulation 
(as well as in the subsequent simulations of more complex knots) 
we used a harmonic representation~\cite{trautwein} in which coordinates
of the monomers are given as polynomials in $\cos(t)$ and $\sin(t)$, 
where $t$ parametrizes the curve.
(This provides a relatively clear visualization of the knot.) 
 Since the hard core and tether potentials do not have an energy scale, 
the temperature $T$ appears in the simulations in the combination
$k_BT/\epsilon_o$, which we will denote as dimensionless temperature $\tilde{T}$.  
All simulations described in this section were performed for $\tilde{T}=1.4$. 
It is customary to represent the strength of the electrostatic potential by 
the {\em Bjerrum length} $\ell_B=e^2/\varepsilon k_BT$. 
(In water at room temperature $\ell_B$=0.7nm.)
In our notation, the Bjerrum length is simply related
to the dimensionless temperature by $\ell_B\equiv a/\tilde{T}$.
(Note that for the moderate values of $N=64$ used in this simulation,
the polymer shape on the right of Fig.~\ref{trefoil_tight} is somewhat `wiggly';
an effect that should disappear for $N\to\infty$ due to the overextensivity 
of the energy.)

Figure~\ref{trefoil_tight} clearly shows that in equilibrium the trefoil
assumes an almost circular shape, with the topological details
concentrated on a very small portion.
(The scale of the right side part of Fig.~\ref{trefoil_tight} has in fact
been reduced by a factor 2 relative to the left figure,
and the actual linear extent of the equilibrated knot is almost twice
its initial size.)
This behavior can be explained by comparing the long and
short-ranged contributions to the Coulomb interaction:
By expanding its radius, the long-range part of the Coulomb energy 
is reduced by a factor of $\delta(N^2/R)\propto N$. 
This comes at the cost of bringing several charges close together in
the tight portion, but the latter energy is independent of $N$, and 
can be easily tolerated for sufficiently long polymers.

\begin{figure}
\includegraphics[width=8cm]{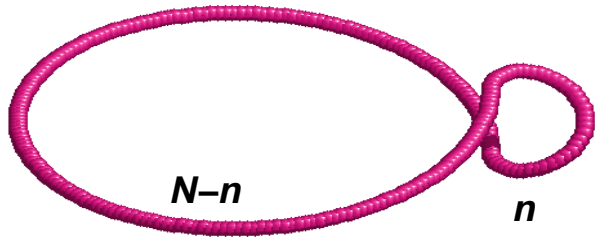}
\caption{\label{simplify}
A closed loop ($N$--monomer polymer)  folded into a shape that 
can be approximately described as two circles consisting of
$N-n$ and $n$ monomers, separated from each other by a distance
of order $n$ monomer sizes.}
\end{figure}

Because of the highly curved portion, Eq.~\ref{Ec} does not apply to tight knots.
For a semi-quantitative understanding of the tension that creates such objects,
consider a simpler example of  an $N$-monomer 
closed chain folded into a shape consisting of a large circle of $N-n$ monomers, 
and a small loop of $n$ monomers, as depicted in Fig.~\ref{simplify}.  
For $n\ll N$, the electrostatic energy can be decomposed as
$E_c(N-n)+E_c(n)+E_i$, where $E_c$ is given in Eq.~\ref{Ec}, 
while $E_i$ is the interaction energy between the small loop and the large circle.
Assuming that the curved strands are separated by a distance of the order $na$, 
the latter is of the order of $2\epsilon_on[\ln(N/n)+c']$, 
where $c'$ is a constant depending on the details of the shape. 
The leading $n$--dependent part of the total energy is then

\begin{equation}
E(N,n)\simeq \epsilon_o\, n\ln\left(\frac{N}{n}\right) ,
\label{En}
\end{equation}
representing a tension that grows logarithmically with the length of the polymer.
This conclusion is not limited to the shape depicted in Fig.~\ref{simplify},
but should apply to any smooth linear curve consisting of two portions 
of very different sizes. 
Equation~\ref{En} thus indicates that from purely electrostatic energy
considerations $n$ should take the smallest possible value,
as indeed happens in the case of a tight knot in Fig.~\ref{trefoil_tight}.

%%%%%%%%%%%%%%%%%%%%%%%%%%%%%%%%%%%%%%%%%%%%%%%%%%%%%%%%%%%%%%%%%%%

%%%%%%%%%%%%%%%%%%%%%%%%%%%%%%%%%%%%%%%%%%%%%%%%%%%%%%%%%%%%%%%%%%%
\begin{figure} 
\begin{center}
\includegraphics[width=6.2cm]{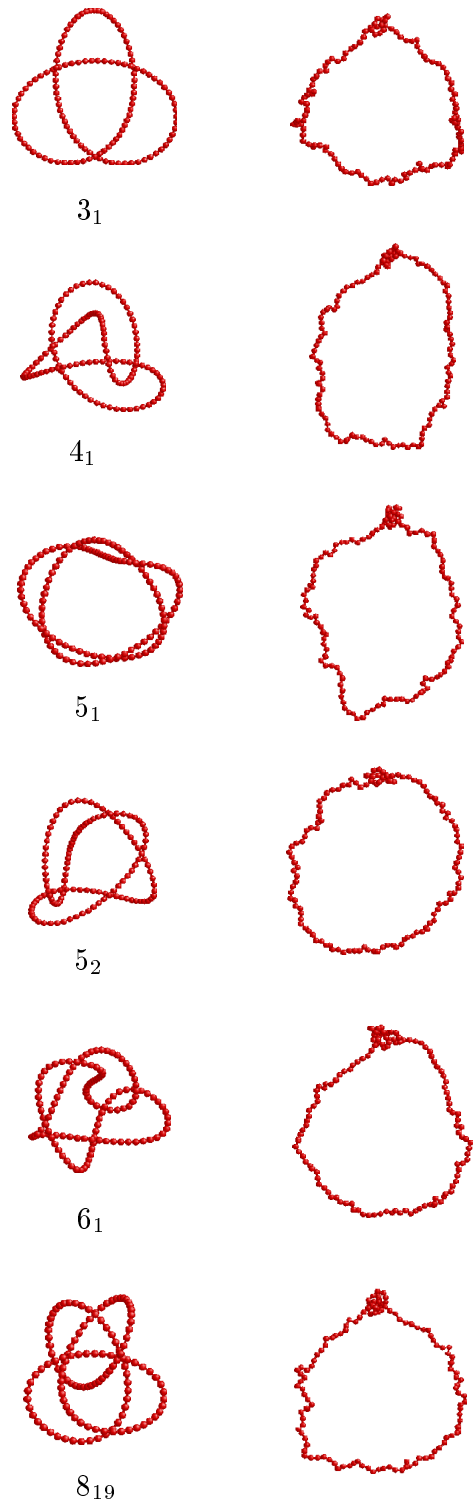}
\caption{\label{prime_knots}The initial (left) and equilibrium (right)
shapes of knots formed by 128-monomer polymers at $\tilde{T}=1.4$
($\ell_{\rm B}=0.7a$).  A selection of prime knots of varying degrees
of complexity is depicted.  (The figures in the right column have been
scaled down.)  The numbers in the left column are the standard
notations for knot types.}
\end{center}
\end{figure} 
%%%%%%%%%%%%%%%%%%%%%%%%%%%%%%%%%%%%%%%%%%%%%%%%%%%%%%%%%%%%%%%%%%%

The tightness observed for a trefoil knot also occurs in more
complicated topologies. Figure~\ref{prime_knots} depicts the results of
equilibration of 128-monomer polymers beginning from a harmonic shape
on the left, to equilibrium shapes (on the right) at $\tilde{T}=1.4$. 
Below each figure we indicate the type of the knot in
the standard notation ${\cal C}_k$, where ${\cal C}$ is the minimal
number of crossings the knot can have in a planar projection \cite{knotbook}. 
Since for a given number of crossings there can exist several different
knots, an additional subscript $k$ labels the standard ordering of these  knots.  
(For ${\cal C}=3$ and 4 there is only
one knot, while for ${\cal C}=8$ there are 21 distinct knots~\cite{knotbook}.)  
Despite the increasing topological complexity of the knots in
Fig.~\ref{prime_knots}, their eventual (collapsed-knot) state is reliably
represented by the semi-quantitative description based on the
energetics of Fig.~\ref{simplify}.

%%%%%%%%%%%%%%%%%%%%%%%%%%%%%%%%%%%%%%%%%%%%%%%%%%%%%%%%%%%%%%%%%%%
\begin{figure} 
\begin{center}
\includegraphics[width=8cm]{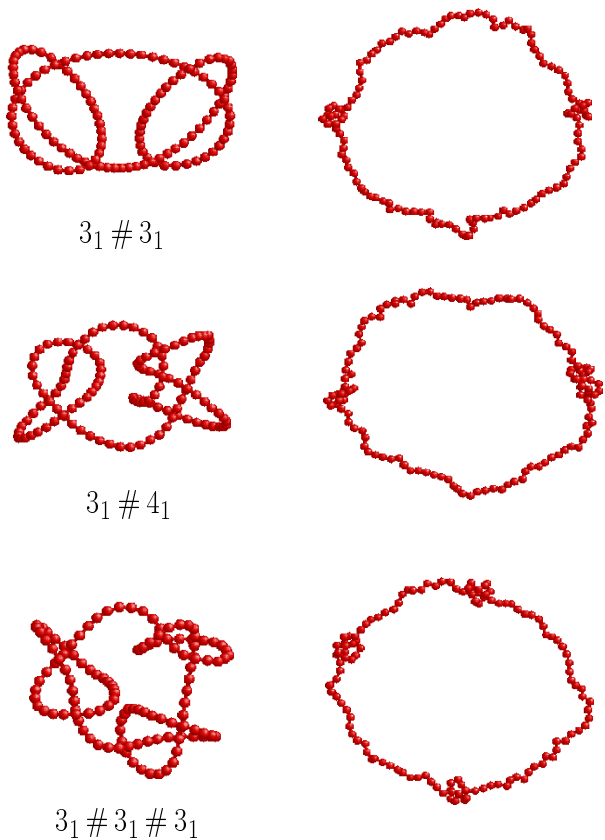}
\caption{``Coulomb factorization'' of composite knots on a 128-monomer
polymer at $\tilde{T}=1.4$. Original (left) and equilibrium (right) 
configurations (scaled down) are shown.}
\label{comp_knot}
\end{center}
\end{figure} 
%%%%%%%%%%%%%%%%%%%%%%%%%%%%%%%%%%%%%%%%%%%%%%%%%%%%%%%%%%%%%%%%%%%

The above arguments indicate the energetic advantage of compressing 
any {\em indivisible} topological constraint into a tight shape
(as opposed to leaving it as an expanded structure). 
However, similar considerations suggest that, {\em if possible},
any concentrated region of charge should split into smaller elements
placed as far as possible from each other. 
Such a reduction is not possible for the {\em prime knots} considered
in Fig.~\ref{prime_knots}, which (by definition)
cannot be separated into several parts connected by a single line. 
In contrast, {\em composite knots} are formed by joining several prime 
factors together, and Fig.~\ref{comp_knot} presents initial and final 
(equilibrium) states of several such knots on 128-monomer polymers. 
%The simulation was performed at $\tilde{T}=1.4$. 
The notation below each knot indicates its constituent prime components. 
The Coulomb interaction clearly ``factorizes'' any composite knot, 
separating its elements as far as possible. 
However, since the typical interaction energies between the prime
factors are only a few $\epsilon_o$,  thermal fluctuations ($\tilde{T}=1.4$)
in the distances between these tight regions are quite pronounced.

\section{Beyond `Ideal' Knots}\label{real}

Many of the results in the previous section are in fact known to knot theorists,
who have investigated long-range repulsive interactions with the
aim of finding a knot-invariant energy~\cite{idealknots,hoidn}. 
The basic question is whether a properly scaled energy of the ground state
configuration (the {\em ideal} state) for certain choices of interaction functions 
can be used as a means of distinguishing different knot types.  
An example of such an interaction is Simon's `minimal distance' between the 
strands function, or a repulsive  $1/r^2$ type interaction~\cite{simon} which produces
symmetric spread out ground states.
In Ref.~\cite{kusner} it was conjectured that minimizing
knot-invariant energies should decompose a knot into prime sub-knots
and simulations with $1/r^2$ interactions support this~\cite{kauffman}.  
Electrostatic interactions do not generate useful knot-invariant energies,
since, in the absence of excluded volume interactions, knots on a 
continuous curve are collapsed to a point~\cite{ohara}, 
providing no (cut-off independent) way of identifying knots.
(Indeed, in the simulations of the previous section knots were tightened into compact
objects whose extent was determined by the monomer size.)
While this conclusion may be disappointing to a knot--theorist, it is encouraging
from the perspective of polymer science, since it is easier to describe the
properties of tight entanglements, without having to worry about their precise topology.
However, this is the case only if we can demonstrate that tight knots survive for
realistic polymers subject to electrostatic interactions in actual solvents.
Accordingly, in this section we shall include additional attributes present in
such situations, and consider the effects of bending rigidity, 
thermal fluctuations, and (most importantly) of a finite screening length. 
In these circumstances the size of the knot can be
significantly larger than in its maximally tight state;
nevertheless, tight knots can still remain.

\subsection{Bending rigidity}\label{ell-p}

Many microscopic aspects of polymers are captured at a mesoscopic 
scale by a curvature energy, describing its resistance to bending.
In a charged polymer one should distinguish between the {\em intrinsic} 
bending rigidity, and an {\it effective} rigidity which includes the 
electrostatic contributions. 
The latter arises because bending a straight segment brings the monomers 
closer and thus increases the Coulomb energy.  
The former can be represented by a length $\ell_p$ at which, in the absence 
of other interactions, the transverse thermal fluctuations of the polymer 
become of the same order as the length-scale itself, 
or at which orientations of the bonds become uncorrelated.
Simple analysis relates $\ell_p$ to the bending rigidity $\kappa$
and temperature by $\kappa\equiv k_BT\ell_p$.  
In charged polymers, $\ell_p$ should be measured in the presence 
of high salt content, so that electrostatic contributions to rigidity are screened out.
It is reasonable that the bending rigidity, rather than monomer size, should
determine the size of a tight knot. 
The energy for bending a segment of length $\ell$, with radius of curvature also 
of order of $\ell$, is $\kappa/\ell$ with a dimensionless shape-dependent prefactor. 
For the shape depicted in Fig.~\ref{simplify}, there is now a bending cost of
$E_b\approx \kappa/na$ which competes with the electrostatic energy in Eq.~\ref{En}.  
By minimizing the sum of these energies, we find that the optimal knot size is
\begin{eqnarray}
\label{nk}
n_k &\approx& \sqrt{\frac{\kappa}{\epsilon_o \ln (N/n_k)}}
\approx \sqrt{\frac{\kappa}{\epsilon_o \ln(N^2\epsilon_o/\kappa)}}\nonumber \\
&=& \sqrt{\frac{\ell_{\rm p}}{\ell_{\rm B} \ln(N^2\ell_{\rm B}/\ell_{\rm p})}} \ ,
\end{eqnarray}
where we have omitted numerical prefactors of order unity. This result
indicates that the knot in stiff polymers of moderate size $N$ can be
as large as $\sqrt{\kappa/\epsilon_o}=\sqrt{\ell_{\rm
p}/\ell_{\rm B}}$, and  becomes compact only for
$N\sim\exp(\kappa/\epsilon_o)$.

\subsection{Thermal Fluctuations}\label{temp}

%%%%%%%%%%%%%%%%%%%%%%%%%%%%%%%%%%%%%%%%%%%%%%%%%%%%%%%%%%%%%%%%%
\begin{figure}[t]
\begin{center}
\includegraphics[width=8cm]{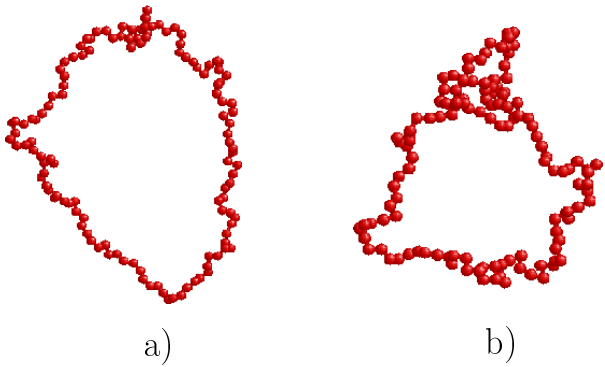}
\caption{Equilibrium configurations of a 128-monomer trefoil knot: 
(a) a tight ($\sim20$ monomer) knot at $\tilde{T}=5$, 
and (b) an expanded ($\sim60$ monomer) knot at $\tilde{T}=10$.}
\label{finite_T_blobs}
\end{center}
\end{figure} 
%%%%%%%%%%%%%%%%%%%%%%%%%%%%%%%%%%%%%%%%%%%%%%%%%%%%%%%%%%%%%%%%%%

At high temperatures, entropic factors (which favor crumpled states)
compete with electrostatic effects. 
While the latter dominate on sufficiently long length-scales, 
at short length-scales fluctuations are important. 
This competition can be visualized by a simple {\em blob} picture~\cite{degennesSC}. 
If a strong external force $f$ is applied to a self-avoiding polymer without 
electrostatic interactions, it is stretched to a linear form. 
This linear object, however, has a finite width $R_b$, and can be regarded as a
chain of blobs of this size.  
On length--scales shorter than the blob size, the external force has negligible effect, 
and we can relate $R_b$ to the number of monomers $N_b$ forming the blob 
via the usual relation for self--avoiding polymers\cite{degennesSC}:
$R_b\approx aN_b^\nu$ with $\nu\approx0.59$. 
Consequently, the linear extent of the entire polymer is approximately $R_b(N/N_b)$. 
If a weak force $f$ is applied to a segment of spatial extent $R_b$, 
that segment is stretched\cite{degennesSC} by an amount $X\approx R_b^2f/k_BT$.
The size of a blob is determined by requirement that $X\approx R_b$,
leading to $R_b\approx k_BT/f$. 
An open charged polymer can also be viewed as a stretched chain formed from such
blobs \cite{degennes76,pfeuty78}, while a ring-polymer is a circle of such blobs.  
The force stretching a blob in an object of this type
is $\epsilon_o a (N_b/R_b)^2\ln(N/N_b)$. 
By substituting this force into the expression for blob size, and solving it, 
we extract the number of monomers in each blob as
\begin{eqnarray}
\label{Nb}
N_b&\approx&\left[\frac{k_BT}{\epsilon_o\ln(N/N_b)}\right]^\frac{1}{2-\nu}
\approx \left[\frac{\tilde{T}}{\ln(N/\tilde{T}^\frac{1}{2-\nu})}\right]^\frac{1}{2-\nu}\nonumber
\\
&=&\left[\frac{a/\ell_{\rm B}}{\ln{N(a/\ell_{\rm B})^\frac{1}{2-\nu}}}\right]^\frac{1}{2-\nu}.
\end{eqnarray}
Of course, the blob picture is meaningful only if $N_b$ is larger than unity.
Thus blobs can appear only for temperatures $\tilde{T}\gg \ln N$;
and for $N=128$ we expect to see the blobs for $\tilde{T}\agt5$.  
Fig.~\ref{finite_T_blobs} depicts equilibrium shapes of a trefoil knot
at $\tilde{T}=5$ and $\tilde{T}=10$, and we see the appearance of 
a wiggly structure in the higher temperature regime. 
At such high temperatures, we expect knots to have a size typical 
of that in a non--charged polymer consisting of $N_b$ monomers. 
The exact size of the knot region in non--charged polymers
in three--dimensional space is not known; simulations suggest that knots are
localized~\cite{katrich,orlandini}, but not compact~\cite{farago}. 
 The size of the blob in Fig.~\ref{finite_T_blobs} is too small for any kind 
of quantitative study, but we clearly see that the knot is no longer maximally compact.

\subsection{Screened Interactions}\label{screen}

A charged polymer in solution is accompanied by neutralizing counterions,
and potentially other charged ions due to added salt.
In general, the effect of these additional ions  on the charged polymer 
is quite complicated, and dependent on the intrinsic stiffness, strength 
of the charge, and valency of counterions~\cite{polyel_pers}. 
However, in many cases the net effect can be approximated by a
screened Coulomb potential $V=(e^2/\varepsilon r)\exp{(- r/\lambda)}$, 
where $\lambda$ is the Debye screening length~\cite{debye}. 
Since the previous arguments for the tightness of charged knots
rely on the long-ranged part of the Coulomb interaction, we may well
question if and when tight knots survive with screened forces.

It is important to realize that Coulomb interactions affect the polymer
on scales much larger than $\lambda$, due to increased bending rigidity.
Curving a straight polymer to a radius $R$ brings its charges closer,
resulting in an extra energy cost of order $(e^2/\varepsilon R)\tilde{\lambda}^2$ 
for screened Coulomb interactions,
where $\tilde{\lambda}\equiv\lambda/a$ is the reduced screening length.
This can be regarded as an effective bending rigidity, which (in the
presence of thermal fluctuations) leads to the
Odijk--Skolnick--Fixman persistence length~\cite{OSF} of
$\ell_c=\lambda^2e^2/(\varepsilon k_BT a^2)=\ell_{\rm B}\tilde{\lambda}^2$. 
The electrostatic persistence length is in general much larger than
the screening length.
In terms of our reduced variables
$\tilde{\ell}_c\equiv\ell_c/a=\tilde{\lambda}^2/\tilde{T}$.  This
expression is valid provided that the length scales considered are
larger than the screening length, and $\tilde{T}< \tilde{\lambda}$.

For very large $\lambda$, comparable to the size of
the polymer, the effects of screening are not very important: E.g.,
Eq.~\ref{En} for electrostatic energy of a knot remains valid if $N$
is replaced by $\tilde{\lambda}$,
and similar replacements should be made in Eq.~\ref{nk} for the knot
size in a stiff polymer, or in Eq.~\ref{Nb} for the blob size. In all
these expressions, the number of monomers enters only in a logarithm,
and, consequently, its replacement by $\tilde{\lambda}$ does not
significantly change the result.  Eq.~\ref{nk} for the optimal knot
size is valid (with $N$ replaced by $\tilde{\lambda}$) only if  the knot
is smaller than the screening length. This condition, $\lambda> an_k$,
leads to the cross-over boundary
\begin{equation}
\lambda > a\sqrt{\frac{\ell_{\rm p}}{\ell_{\rm B}}},
\end{equation}
which is equivalent to $\ell_{\rm c}> \ell_{\rm p}$. 
We thus conclude that a tight knot can exist only when 
the overall bending rigidity is dominated by electrostatic contributions.  
For smaller values of $\lambda$, the short range repulsion can no
longer bend the knot into a tight shape.

%%%%%%%%%%%%%%%%%%%%%%%%%%%%%%%%%%%%%%%%%%%%%%%%%%%%%%%%%%%%%%%%%
\begin{figure}[t]
\begin{center}
\includegraphics[width=8cm]{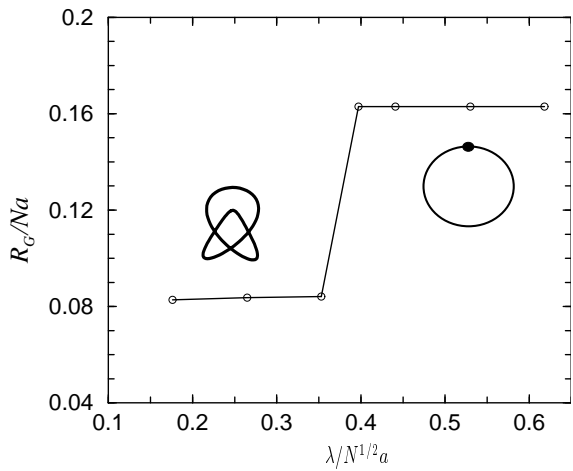}
\caption{Radius of gyration, $R_{\rm g}$, of the ground state
configuration of a trefoil knot,
as a function of the screening length $\lambda$ for a 128-monomer polymer.
$R_{\rm G}$ has been normalized by the length of the polymer  $Na$.}
\label{Rg}
\end{center}
\end{figure} 
%%%%%%%%%%%%%%%%%%%%%%%%%%%%%%%%%%%%%%%%%%%%%%%%%%%%%%%%%%%%%%%%%%

Note that the analysis leading to Eq.~\ref{nk} only demonstrates the
local stability of a tight knot.  
The global energy minimum could still occur for a spread out configuration. 
To decide on the latter requires estimates of the energy difference between 
the two configurations, and depends on microscopic details,
as well as the length of the polymer.  
A circle with a tight knot, and the spread out knotted shape, both
have a bending energy (at large scales) of the order of $k_{\rm B}T \ell_{\rm c}/R$.
Since the circular shape has a larger radius, it has a lower energy,
the energy difference scaling as $k_{\rm B}T \ell_{\rm c}/(Na)$, if both radii are
proportional to the polymer length.
The tight knot in the former has an additional local energy cost, which is
of the order of $k_BT(\ell_{\rm B}/a)$ (possibly with logarithmic corrections),
but independent of $N$. 
Thus, we expect the configuration with a spread-out knot to have 
a lower energy only
for $\ell_{\rm c}/N<\ell_{\rm B}$, i.e. for screening lengths
\begin{equation}
\lambda\leq \lambda_c\approx a \sqrt{N}.
\end{equation}
Note that the limiting value of $\lambda$ still corresponds to a persistence 
length of the order of the extended polymer, i.e. the polymer shape is
determined by energy considerations, and thermal fluctuations have little effect, at this point.
We verified this conclusion by numerically determining the shape
of the trefoil that minimizes the screened Coulomb interactions. 
Figure~\ref{Rg} shows the radius of gyration as a function of the screening length. 
For screening lengths larger than $\lambda_c\sim 0.4 aN^{1/2}$, 
the knot switches from a loose to a tight configuration.

Let us briefly explore the possibility of tight knots in nucleic acids.
Double stranded DNA has a bare persistence length of $\ell_{p}\sim 50$nm,
which is much larger than typical screening lengths, and consequently
is not likely to incorporate any knots tightened by Coulomb interactions.
However, measurements on {\em single stranded} DNA in high salt
concentrations~\cite{tinland} suggest a much smaller intrinsic $\ell_{p}\sim 1$nm,
and presumably a similar (or smaller) value applies to single stranded RNA.
Tight knots should then occur for single stranded nucleic acids for
reasonable screening lengths of the order  $\lambda\sim 10$nm.
This could for example be relevant to the experiments of Ref.~\cite{wang},
where artificial knots in single stranded RNA were used to demonstrate
the existence of a topology changing enzyme.
Knotted polymers are often distinguished from unknotted ones by 
electrophoresis~\cite{wang}.
However, if the knot is tight, the knotted polymer may have an electrophoretic
mobility close to that of a ring polymer, making such detection problematic.

\section{Tight Knots and Dynamics}\label{dynamics}

Tight knots are created whenever a polymer is under tension;
the source of tension need not be long-range repulsions.
For example, it has been argued that tight molecular knots appear in polymer 
systems undergoing crystallization, as crystallization at one point may create 
tension in other parts of the chain~\cite{degennes}.  
Polymers in a strong shear flow are also subject to 
tension~\cite{degennes,degennes_book}, and may even undergo a
coil-stretch transition as a result~\cite{degennes2}.
It is plausible that stretching could tighten loose knots in the chain.
Once created, such molecular knots should be quite stable and thus
account for long-time memory effects observed in polymer melts~\cite{degennes}.  
However, molecular dynamics simulations suggest that once the tension is
removed a tight knot opens up in a short time~\cite{mansfield}.
Without being systematic, here we examine a couple of dynamical issues
pertaining to charged tight knots, namely their creation in a high temperature
quench, and their relaxation by diffusion along the chain.

\subsection{Tightening by Quench}

\begin{figure}
\begin{center}
\includegraphics[width=7cm]{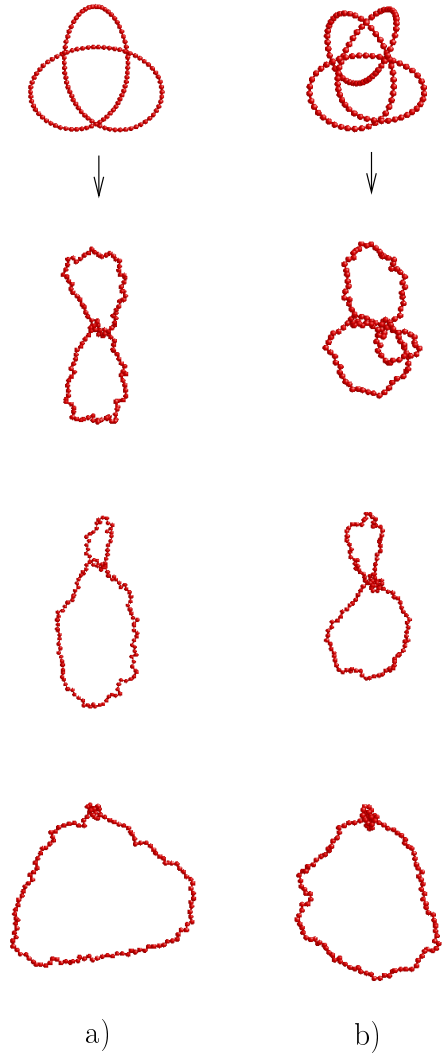}
\caption{\label{tight}Time evolution (using Monte Carlo dynamics)
of (a) $3_1$ and  (b) $8_{19}$ knots, from the initial (harmonic)
geometry (top) through an intermediate state when the knot ``strangles''
the loop close its middle, and to a final state (bottom) when the
knot is localized. A similar sequence takes place
for all other prime knots in the simulations of Fig.~\ref{prime_knots}.}
\end{center}
\end{figure}

It is quite likely that when topological entanglements are first formed,
e.g. in the process of cyclization of a polymer,
they are spread out over the whole chain.
Subsequent tightening then occurs upon increasing tension.
In the case of charged knots, this process is illustrated in Fig.~\ref{tight}.
Here, the initial configurations are the spread out harmonic 
representations, which soon evolve into loops separated by tight elements.
The relaxation process then slows down as one of
the loops grows at the expense of the others.
A universal last stage is the appearance of a structure reminiscent
of Fig.~\ref{simplify}, with two loops separated by a tight `slip-link.'
We observed the same sequence in simulations where the initial 
configurations was an equilibrated (random walk) knot.  
The formation of the two loops separated by a slip-link was again
relatively fast, and the rate limiting step was the sliding of one loop
through the tightly packed  monomers at the slip-link.

%%%%%%%%%%%%%%%%%%%%%%%%%%%%%%%%%%%%%%%
%%%%%%%%%%%%%%%%%%%%%%%%%%%%%%%%%%%%%%%%%%%%%%%%%%%%%%%%%%%%%%%%%%
%%%%%%%%%%%%%%%%%%%%%%%%%%%%%%%%%%%%%%%%%%%%%%%%%%%%%%%%%%%%%%%%%%%%
\begin{figure}
\begin{center}
\includegraphics[width=8cm]{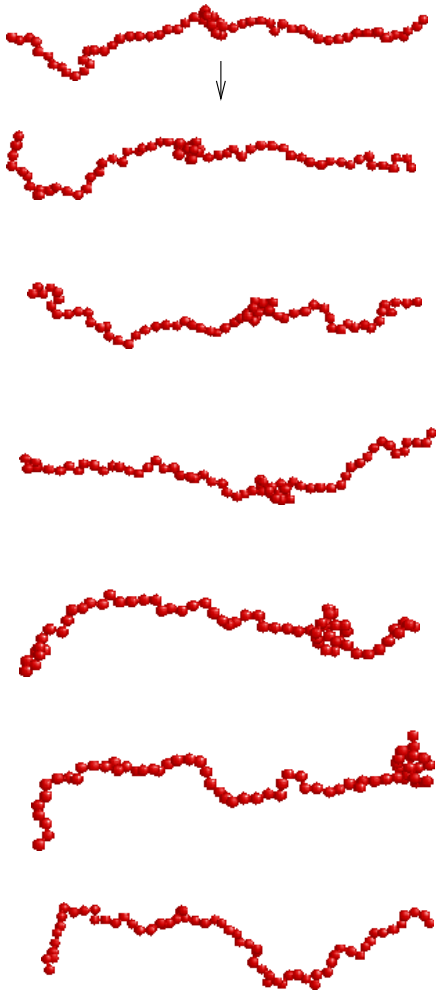}
\caption{\label{unknot_lin}Unknotting of a charged polymer with
 $N=64$ monomers and
unscreened Coulomb interaction of strength $\tilde{T}=1.4$ 
($\ell_{\rm B}=0.7a$). The initial configuration is a
tight knot in the middle of the chain. Rather than open up gradually,
the knot slides along the polymer and remains localized until it
reaches the end.}
\end{center}
\end{figure}
%%%%%%%%%%%%%%%%%%%%%%%%%%%%%%%%%%%%%%%%%%%%%%%%%%%%%%%%%%%%%%%%%%%% 

\subsection{Diffusion of Tight Knots}

As demonstrated in the previous situation, tight knots slow down the
relaxation of the polymer to its eventual equilibrium state.
Here we study such relaxation more explicitly for a knot
in an open charged polymer.
In this case there is no topological constraint, and the polymer
is expected to unknot to achieve its equilibrium state.
Does a tight knot in an open chain relax by becoming loose
and opening up, or by sliding (diffusing) to one end.
As demonstrated in Fig.~\ref{unknot_lin}, the latter is the case:
The initial configuration (in a chain of $N=64$ monomers with unscreened
interaction) remains tight, indicating that the stretching force from the 
monomers at the ends of the chain is larger than from those forming the knot.
In the simulation, the  knot's position fluctuates for some time in the middle, 
before moving to one direction.  
The eventual unknotting occurs when the diffusing tight knot
reaches the end of the polymer.  

A tight knot in the middle of an open chain is in a meta-stable state.
We can estimate a potential energy for the tight knot by considering 
a charge $Q=n e$ along a charged chain of $N$ monomers.
The Coulomb energy then depends on the position of this charge
$N_1$, as $E=k_{\rm B}T(\ell_{\rm B}/a) n \ln[N_1(N-N_1)]$.  
This energy is minimal when the charge $Q$ is
at either endpoint of the polymer, i.e. for $N_1=0$ or $N$. 
Note that the force pushing the extra charge towards the end scales
with $\ell_{\rm B}$, and we may naively expect that the
resulting relaxation becomes faster as the Coulomb energy is increased.
In fact, the opposite occurs for charge knots, with relaxation slowing 
down as Coulomb interactions become more dominant.
The reason is that increased charging energy leads to a higher
tension and more closely packed monomers in the knot.
Any motion of the knot requires some internal rearrangements
of these monomers, accompanied by pulling in some monomers from
the straight portions of the chain.
This necessitates overcoming an energy barrier of $\sim \ell_{\rm B}\ln{N}$,
and consequently higher charged knots are tight and harder to move.
Since rearrangements require a large activation energy, the knot
remains stuck in position. 
This is quite similar to what happens to a knot in a polymer under 
strong tension~\cite{degennes}.

%%%%%%%%%%%%%%%%%%%%%%%%%%%%%%%%%%%%%%%%%%%%%%%%%%%%%%%%%%%%%%%%%
\begin{figure}
\begin{center}
\includegraphics[width=8cm]{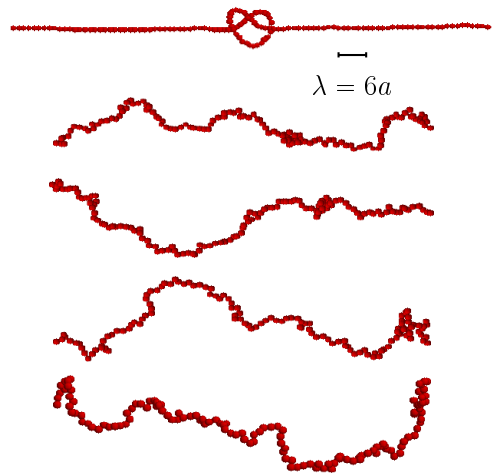}
\caption{\label{unknot_scr}
Monte Carlo dynamics of a tight knot  in a chain with $N=128$
monomers and screened interactions.  The screening length is
$\lambda=6a$, roughly the size of the knot. The knot shows no sign of
opening up, it remains tight till it reaches the end of the polymer. }
\end{center}
\end{figure} 
%%%%%%%%%%%%%%%%%%%%%%%%%%%%%%%%%%%%%%%%%%%%%%%%%%%%%%%%%%%%%%%%%%

While with unscreened Coulomb interactions the tight knot feels a 
potential that drives it to one end, there is no such force when the 
interactions are screened (unless the distance between the knot and 
the endpoint of the polymer is of the order of the screening length).  
The energy barrier preventing the loosening of the knot is also
finite in this case.
The resulting dynamics for a chain of 128 monomers with a screening
length of $\lambda=6a$ is demonstrated in Fig.~\ref{unknot_scr};
despite the screening the knot remains tight until it diffuses to one end.
The characteristic time scales for the relaxation of the knot can be 
estimated as follows.
The time for diffusion over a distance $Na$ scales as $a^2N^2/D_{knot}$,
with the knot diffusion coefficient behaving as 
$D_{knot}\propto D\exp{(-{\cal E}_D/k_{\rm B}T)}$. 
Here, $D$ is the diffusion constant for a single monomer, while the activation 
energy for local rearrangements necessary for motion of the tight region is 
roughly ${\cal E}_D\approx k_{\rm B}T(\ell_{\rm B}/a) \log(\lambda/a)$.
There is also the possibility that the knot becomes loose, escaping the
local minimum of the tight configuration.
The energy barrier for the latter is
${\cal E}_b\approx k_BT\ell_{\rm B}\lambda/a^2$, with a corresponding
time scale of   $\tau \approx (a^2/D)\exp{({\cal E}_b/k_{\rm B}T)}$.
In time $\tau$, the knot can diffuse a distance $L \approx \sqrt{D_{knot} \tau}$. 
We thus estimate a ``processivity length'' over which a tight knot diffuses, 
before opening up, by
\begin{equation}
L_{\rm p}\propto a \exp{\left( C\; \ell_{\rm B}\lambda/a^2\right)},
\end{equation}
where $C$ is a constant of order unity.  The processivity length increases
strongly with the screening length $\lambda$ and quickly reaches a
macroscopic length, indicating that the relaxation of a tight knot
will be by diffusion along the chain, even for very long chains. 
Also note that $L_{\rm p}$ is in general much larger than the 
electrostatic persistence length which only grows quadratically 
with the screening length ($\ell_{\rm c}\approx \ell_{\rm B}\lambda/a^2$).

\section{Discussion}\label{discussion}
We have shown that long-ranged Coulomb forces 
generate a tension that tightens topological constraints
into dense localized regions, leaving the rest of the polymer unentangled.
For knots on ring polymers, we confirm the ``factorization" of 
composite knots into their prime components.
Tight knots remain, even when the Coulomb interaction is screened,
as long as the electrostatic contributions dominate the rigidity of the polymer.
Once formed, tight knots drastically slow down the equilibration 
of the polymer (or polymer solution), as they typically relax by
diffusion along the backbone.
If the Coulomb interactions are strong enough, the knot is pulled
so tight that it is unable to diffuse, and its position appears frozen.  
This is different from uncharged polymers where molecular dynamics 
simulations in ref.~\cite{mansfield} find that tight knots in short 
uncharged polymers open up rapidly. 
Our results predict that tight knots in polyelectrolytes can be
very stable and cause long relaxation times.
While we have focused on single polymers, it is natural to
speculate about similar behavior in solutions of many chains.
It is indeed quite likely that {\em inter-chain} entanglements
are also tightened in polyelectrolyte solutions and gels.

Additional consequences of tight knots are in their influence on
mobility (electrophoresis), and on the mechanical strength of polymers.
It has been shown recently by direct measurement on DNA and actin 
filaments that knots significantly weaken the strand~\cite{arai}.
Similarly,   molecular dynamics simulations of  knotted polyethylene
chains also find that the strands becomes weaker, and typically break
at the entrance point where the straight segment ends and the tight knot
begins~\cite{saitta}.  Single stranded DNA is relatively fragile 
and  sometimes  breaks during electrophoresis or when subject to flow; 
tight knots may well be responsible for this phenomenon.

\section{Acknowledgements}
We thank Ralf Metzler and Andreas Hanke for helpful discussions. 
This work was supported by the National Science Foundation grants 
DMR-01-18213 and PHY99-07949, and by US-Israel Binational Science 
Foundation grant 1999-007. P.G.D. acknowledges support from the 
Research Council of Norway.

\end{document}